\documentclass[a4paper]{article}

\usepackage{INTERSPEECH2018}
\usepackage{multirow}

\title{Syllable-Based Sequence-to-Sequence Speech Recognition with the Transformer in Mandarin Chinese}
\name{Shiyu Zhou$^1$$^,$$^2$, Linhao Dong$^1$$^,$$^2$, Shuang Xu$^1$, Bo Xu$^1$ \thanks{The research work is supported by the National Key Research and Development Program of China under Grant No. 2016YFB1001404.}}
\address{
  $^1$Institute of Automation, Chinese Academy of Sciences \\
  $^2$University of Chinese Academy of Sciences }
\email{\{zhoushiyu2013, donglinhao2015, shuang.xu, xubo\}@ia.ac.cn}

\begin{document}

\maketitle
\begin{abstract}
Sequence-to-sequence attention-based models have recently shown very promising results on automatic speech recognition (ASR) tasks, which integrate an acoustic, pronunciation and language model into a single neural network. In these models, the Transformer, a new sequence-to-sequence attention-based model relying entirely on self-attention without using RNNs or convolutions, achieves a new single-model state-of-the-art BLEU on neural machine translation (NMT) tasks. Since the outstanding performance of the Transformer, we extend it to speech and concentrate on it as the basic architecture of sequence-to-sequence attention-based model on Mandarin Chinese ASR tasks. Furthermore, we investigate a comparison between syllable based model and context-independent phoneme (CI-phoneme) based model with the Transformer in Mandarin Chinese.
Additionally, a greedy cascading decoder with the Transformer is proposed  for mapping CI-phoneme sequences and syllable sequences into word sequences. Experiments on HKUST datasets demonstrate that syllable based model with the Transformer performs better than CI-phoneme based counterpart, and achieves a character error rate (CER) of \emph{$28.77\%$}, which is competitive to the state-of-the-art CER of $28.0\%$ by the joint CTC-attention based encoder-decoder network.
\end{abstract}
\noindent\textbf{Index Terms}: ASR, multi-head attention, syllable based acoustic modeling, sequence-to-sequence

\section{Introduction}

Experts have shown significant interest in the area of sequence-to-sequence modeling with attention \cite{chiu2017state, chorowski2015attention, bahdanau2016end, chan2015listen} on ASR tasks in recent years. Sequence-to-sequence attention-based models integrate separate acoustic, pronunciation and language models of a conventional ASR system into a single neural network \cite{prabhavalkar2017analysis} and do not make the conditional independence assumptions as in standard hidden Markov based model \cite{bourlard2012connectionist}.

Sequence-to-sequence attention-based models are commonly comprised of an \emph{encoder}, which consists of multiple recurrent neural network (RNN) layers that model the acoustics, and a \emph{decoder}, which consists of one or more RNN layers that predict the output sub-word sequence. An \emph{attention} layer acts as the interface between the encoder and the decoder: it selects frames in the encoder representation that the decoder should attend to in order to predict the next sub-word unit \cite{prabhavalkar2017analysis}. However, RNNs maintain a hidden state of the entire past that prevents parallel computation within a sequence. In order to reduce sequential computation, the model architecture of the Transformer has been proposed in \cite{vaswani2017attention}. This model architecture eschews recurrence and instead relies entirely on an attention mechanism to draw global dependencies between input and output, which allows for significantly more parallelization and achieves a new single-model state-of-the-art BLEU on NMT tasks \cite{vaswani2017attention}. Since the outstanding performance of the Transformer, this paper focuses on it as the basic architecture of sequence-to-sequence attention-based model on Mandarin Chinese ASR tasks.

Recently various modeling units of sequence-to-sequence attention-based models have been studied on English ASR tasks, such as graphemes, CI-phonemes, context-dependent phonemes and word piece models \cite{chiu2017state, prabhavalkar2017analysis, prabhavalkar2017comparison}. However, few related works have been explored by sequence-to-sequence attention-based models on Mandarin Chinese ASR tasks.
As we know, Mandarin Chinese is a syllable-based language and syllables are their logical unit of pronunciation.
These syllables have a fixed number (around 1400 pinyins with tones are used in this work) and each written character corresponds to a syllable.
In addition, syllables are a longer linguistic unit, which reduces the difficulty of syllable choices in the decoder by sequence-to-sequence attention-based models. Moreover, syllables have the advantage of avoiding out-of-vocabulary (OOV) problem.

Due to these advantages of syllables, we are concerned with syllables as the modeling unit in this paper and investigate a comparison between CI-phoneme based model and syllable based model with the Transformer on Mandarin Chinese ASR tasks. Moreover, Since we investigate the comparison between CI-phonemes and syllables, these CI-phoneme sequences or syllable sequences from the Transformer have to be converted into word sequences for the performance comparison in terms of CER. The conversion from CI-phoneme sequences or syllable sequences to word sequences can be regarded as a sequence-to-sequence task, which is modeled by the Transformer in this paper. Then we propose a greedy cascading decoder with the Transformer to
maximize the posterior probability $Pr(W|X)$ approximately. Experiments on HKUST datasets reveal that the Transformer performs very well on Mandarin Chinese ASR tasks. Moreover, we experimentally confirm that syllable based model with the Transformer can outperform CI-phoneme based counterpart, and achieve a CER of \emph{$28.77\%$}, which is competitive to the state-of-the-art CER of \emph{$28.0\%$} by the joint CTC-attention based encoder-decoder network \cite{hori2017advances}.

The rest of the paper is organized as follows. After an overview of the related work in Section \ref{label_related work}, Section \ref{label_system_overview} describes the proposed method in detail. we then show experimental results in Section \ref{label_experiment} and conclude this work in Section \ref{label_conclusions}.

\section{Related work}
\label{label_related work}

Sequence-to-sequence attention-based models have shown very encouraging results on English ASR tasks \cite{chiu2017state, prabhavalkar2017comparison, zhang2017very}. However, it is quite difficult to apply it to Mandarin Chinese ASR tasks. In \cite{chan2016online}, Chan et al. proposed Character-Pinyin sequence-to-sequence attention-based model on Mandarin Chinese ASR tasks. The Pinyin information was only used during training for improving the performance of the character model. Instead of using joint Character-Pinyin model, \cite{shanattention} directly used Chinese characters as network output by mapping the one-hot character representation to an embedding vector via a neural network layer.

In this paper, we are concerned with syllables as the modeling unit. Acoustic models using syllables as the modeling unit have been investigated for a long time \cite{qu2017syllable, ganapathiraju2001syllable, wu2007context}. Ganapathiraju et al. have first shown that syllable based acoustic models can outperform context dependent phone based acoustic models with GMM \cite{ganapathiraju2001syllable}. Wu et al. experimented on syllable based context dependent Chinese acoustic model and discovered that context dependent syllable based acoustic models can show promising performance \cite{wu2007context}.
Qu et al. \cite{qu2017syllable} explored the CTC-SMBR-LSTM using syllables as outputs and verified that syllable based CTC model can perform better than CI-phoneme based CTC model on Mandarin Chinese ASR tasks. Inspired by \cite{qu2017syllable}, we extend their work from CTC based models to sequence-to-sequence attention-based models.

Using syllables as the modeling unit, it is natural to consider the conversion from Chinese syllable sequences to Chinese word sequences as a task of labelling unsegmented sequence data. Liu et al. \cite{liu2015chinese} proposed RNN based supervised sequence labelling method with CTC algorithm to achieve a direct conversion from syllable sequences to word sequences.

\section{System overview}
\label{label_system_overview}

\subsection{Transformer model}
\label{label_transformer_model}

The Transformer model architecture is the same as sequence-to-sequence attention-based models except relying entirely on self-attention and position-wise, fully connected layers for both the encoder and decoder \cite{vaswani2017attention}. The encoder maps an input sequence of symbol representations \textbf{x} = $\left( x_1, ..., x_n \right)$ to a sequence of continuous representations \textbf{z} = $\left( z_1, ..., z_n \right)$. Given \textbf{z}, the decoder then generates an output sequence \textbf{y} = $\left( y_1, ..., y_m \right)$ of symbols one element at a time.

\subsubsection{Multi-head attention}
\label{label_multi_head_attention}

An attention function maps a query and a set of key-value pairs to an output, where the query, keys, values, and output are all vectors. The output is computed as a weighted sum of the values, where the weight assigned to each value is computed by a compatibility function of the query with the corresponding key \cite{vaswani2017attention}. Scaled dot-product attention is adopted as the basic attention function in the Transformer, which describes (\ref{eq:eq_scaled_dot_product_atttention:01}):
      \begin{eqnarray}
        \label{eq:eq_scaled_dot_product_atttention:01}
        {Attention(Q, K, V)} = {softmax}{\left(\frac{QK^\mathrm{T}}{\sqrt{d_k}}\right)}V
     \end{eqnarray}
Where the dimension of query Q and key K are the same, which are \emph{d$_{k}$}, and dimension of value V is \emph{d$_{v}$}.

Instead of performing a single attention function, the Transformer employs the multi-head attention (MHA) which projects the queries, keys and values $h$ times with different, learned linear projections to $d_k$, $d_k$ and $d_v$ dimensions. On each of these projected versions of queries, keys and values, the basic attention function is performed in parallel, yielding $d_v$-dimensional output values. These are concatenated and projected again, resulting in the final values. The equations can be represented as follows \cite{vaswani2017attention}:
      \begin{eqnarray}
        \label{eq:eq_multi_head_attention:01}
        {MultiHead(Q, K, V)} = {Concat}{\left({head_1}, ..., head_h\right)}W^O \\
        where\
        \label{eq:eq_attention:02}
        {head_i} = {Attention}{\left({QW_i^Q}, {KW_i^K}, {VW_i^V}\right)}
     \end{eqnarray}
Where the projections are parameter matrices $W_i^Q\in{\mathbb{R}^{{d_{model}}\times{d_k}}}$, $W_i^K\in{\mathbb{R}^{{d_{model}}\times{d_k}}}$, $W_i^V\in{\mathbb{R}^{{d_{model}}\times{d_v}}}$, $W^O\in{\mathbb{R}^{{hd_v}\times{d_{model}}}}$, $h$ is the number of heads, and $d_{model}$ is the model dimension.

MHA behaves like ensembles of relatively small attentions to allow the model to jointly attend to information from different representation subspaces at different positions, which is beneficial to learn complicated alignments between the encoder and decoder.

      \begin{figure}[t]
        \centering
        \includegraphics[width=0.8\linewidth]{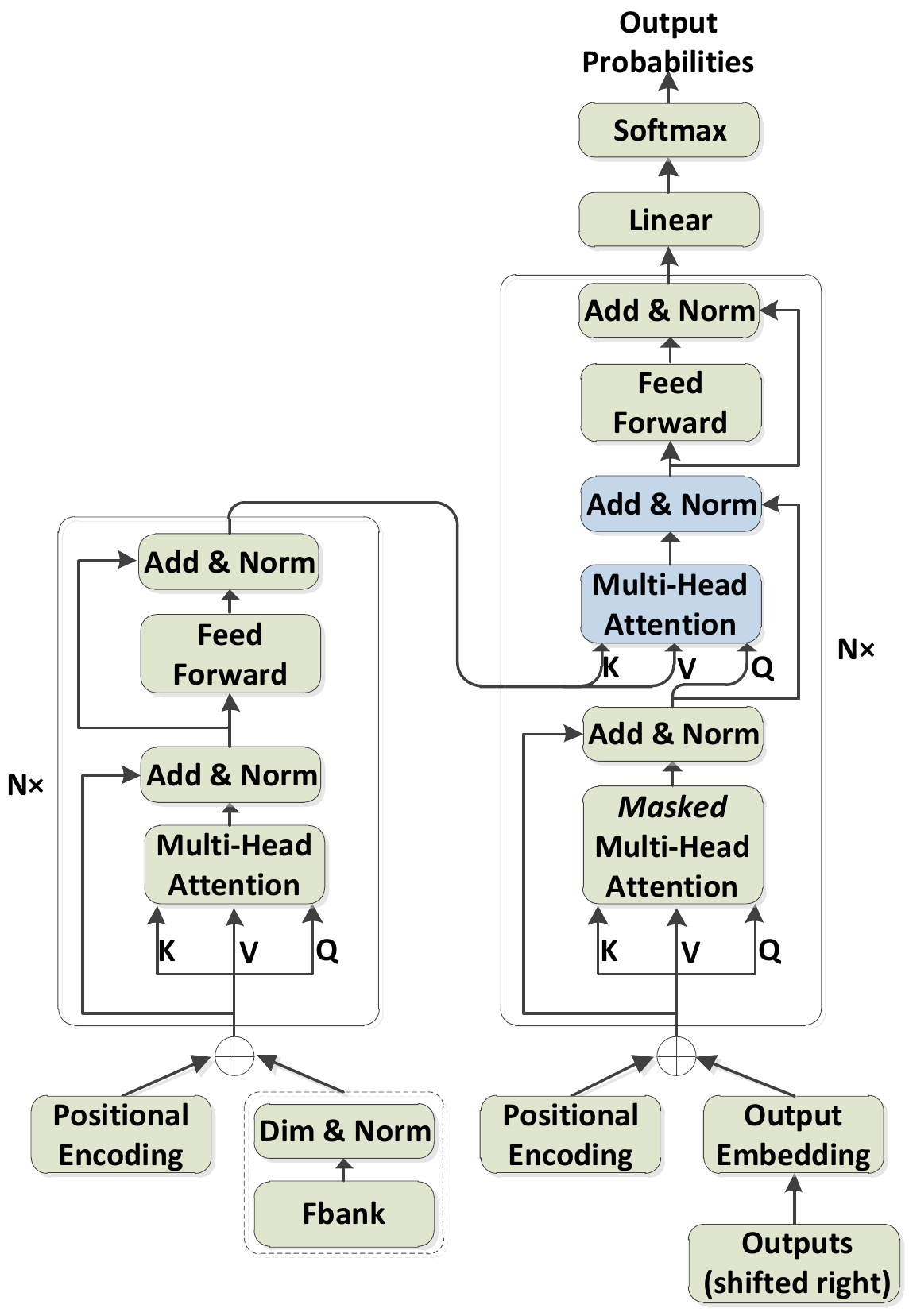}
        \caption{{\it The architecture of the ASR Transformer.}}
        \label{fig:fig_transformer}
      \end{figure}

\subsubsection{Transformer model architecture}

\label{label_transformer_model_architecture}
The architecture of the ASR Transformer is shown in Figure~\ref{fig:fig_transformer}, which stacks MHA and position-wise, fully connected layers for both the encode and decoder. The encoder is composed of a stack of $N$ identical layers. Each layer has two sub-layers. The first is a MHA, and the second is a position-wise fully connected feed-forward network. Residual connections are employed around each of the two sub-layers, followed by a layer normalization. The decoder is similar to the encoder except inserting a third sub-layer to perform a MHA over the output of the encoder stack. To prevent leftward information flow and preserve the auto-regressive property in the decoder, the self-attention sub-layers in the decoder mask out all values corresponding to illegal connections. In addition, positional encodings \cite{vaswani2017attention} are added to the input at the bottoms of these encoder and decoder stacks, which inject some information about the relative or absolute position in the sequence to make use of the order of the sequence.

Since our ASR experiments use 80-dimensional log-Mel filterbank features, we explore a linear transformation with a layer normalization to convert the input dimension to the model dimension $d_{model}$ for dimension matching, which is marked out by a dotted line in Figure~\ref{fig:fig_transformer}.

\subsection{Greedy cascading decoder with the Transformer}
\label{label_decoder}
As syllables and CI-phonemes are investigated in this paper, the CI-phoneme sequences or syllable sequences have to be converted into word sequences using a lexicon during beam-search decoding.

The speech recognition problem can be defined as the problem of finding word sequence W that maximizes posterior probability $Pr(W|X)$ given observation $X$, and can transform as follows \cite{kanda2016maximum}.
     \begin{eqnarray}
        \label{eq:eq_MAP:04}
        \tilde{W} &=& {\operatorname*{argmax}\limits_{W}}\ {Pr}{\left( W | X \right)} \\
        &=& {\operatorname*{argmax}\limits_{W}}{\sum_{s}}\ {Pr}{\left( W | s \right)}{{Pr}{\left( s | X \right)}} \\
        \label{eq:eq_MAP:06}
        &\approx& {\operatorname*{argmax}\limits_{W}}\ {Pr}{\left( W | s \right)}{{Pr}{\left( s | X \right)}}
     \end{eqnarray}
Here, $Pr(s|X)$ is the probability from observation X to sub-word unit sequence $s$, $Pr(W|s)$ is the the probability from sub-word unit sequence $s$ to word sequence $W$.

According to equation~(\ref{eq:eq_MAP:06}), we propose that both $Pr(s|X)$ and $Pr(W|s)$ can be regarded as sequence-to-sequence transformations, which can be modeled by sequence-to-sequence attention-based models, specifically the Transformer is used in the paper.

Then, the greedy cascading decoder with the Transformer is proposed to directly estimate equation~(\ref{eq:eq_MAP:06}). First, the best sub-word unit sequence $s$ is calculated by the Transformer from observation $X$ to sub-word unit sequence with beam size $\beta$. And then, the best word sequence $W$ is chosen by the Transformer from sub-word unit sequence to word sequence with beam size $\gamma$. Through cascading two sequence-to-sequence attention-based models, we assume that equation~(\ref{eq:eq_MAP:06}) can be approximated.

In this work we employ $\beta=13$ and $\gamma=6$.





\section{Experiment}
\label{label_experiment}

\subsection{Data}
The HKUST corpus (LDC2005S15, LDC2005T32), a corpus of Mandarin Chinese conversational telephone speech, is collected and transcribed by Hong Kong University of Science and Technology (HKUST) \cite{liu2006hkust}, which contains 150-hour speech, and 873 calls in the training set and 24 calls in the test set. All experiments are conducted using 80-dimensional log-Mel filterbank features, computed with a 25ms window and shifted every 10ms. The features are normalized via mean subtraction and variance normalization on the speaker basis. Similar to \cite{sak2015fast, kannan2017analysis}, at the current frame $t$, these features are stacked with 3 frames to the left and downsampled to a 30ms frame rate.

\subsection{Training}

We perform our experiments on the \emph{base model} and \emph{big model} (i.e. D512-H8 and D1024-H16 respectively) of the Transformer from \cite{vaswani2017attention}. The basic architecture of these two models is the same but different parameters setting. Table~\ref{tab:paramters} lists the experimental parameters between these two models. The Adam algorithm \cite{kingma2014adam} with gradient clipping and warmup is used for optimization. During training, label smoothing of value $\epsilon_{ls}=0.1$ is employed \cite{szegedy2016rethinking}.
        \begin{table}[th]
        \caption{\label{tab:paramters} {\it Experimental parameters configuration.}}
        \vspace{2mm}
        \centerline{
          \begin{tabular}{|c|c|c|c|c|c|c|}
            \hline
              {model}   & $N$  &    $d_{model}$ & $h$ & $d_k$ & $d_v$   & $warmup$  \\
            \hline
            D512-H8 &  $6$ &   $512$  & $8$ & $64$  &  $64$  & $4000\ steps$ \\
            \hline
            D1024-H16 & $6$ &   $1024$ & $16$ & $64$ &  $64$  & $12000\ steps$ \\
            \hline
          \end{tabular}
        }
        \end{table}

First, for the Transformer from observation $X$ to sub-word unit sequence, $118$ CI-phonemes without silence (phonemes with tones) are employed in the CI-phoneme based experiments and $1384$ syllables (pinyins with tones) in the syllable based experiments. Extra tokens (i.e. an unknown token (\textless UNK\textgreater), a padding token (\textless PAD\textgreater), and sentence start and end tokens (\textless S\textgreater/\textless \textbackslash S\textgreater)) are appended to the outputs, making the total number of outputs $122$ and $1388$ respectively in the CI-phoneme based model and syllable based model. Second, for the Transformer from sub-word unit sequence to word sequence, we collect all words from the training data together with appended extra tokens and the total number of outputs is $28444$. In our experiments, we only train the Transformer from sub-word unit sequence to word sequence by the base model.

Standard tied-state cross-word triphone GMM-HMMs are first trained with maximum likelihood estimation to generate CI-phoneme alignments on training set and test set for handling multiple pronunciations of the same word in Mandarin Chinese. we then generate syllable alignments through these CI-phoneme alignments according to the lexicon. Finally, we proceed to train the Transformer with these alignments.

In order to verify the effectiveness of the greedy cascading decoder proposed in this paper, the CI-phoneme and syllable alignments on test data are converted into word sequences using the trained $Pr(W|s)$ models. We can get a CER of $4.70\%$ on the CI-phoneme based model and $4.15\%$ on the syllable based model respectively, which are the lower bounds of our experiments. If sub-word unit sequences, calculated by the Transformer from observation $X$ to sub-word unit sequence $s$, can approximate to these corresponding alignments, our experimental results can approach the lower bounds using the greedy cascading decoder.

\begin{figure}[t]
            \centering
            \includegraphics[width=5.0cm]{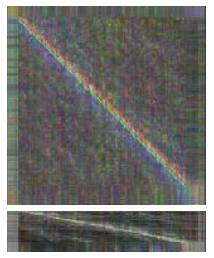}
            \caption{{\it Self-attention (top) of encoder-encoder that both x-axis and y-axis represent input frames. Vanilla attention (bottom) of encoder-decoder that the x-axis represents input frames and y-axis corresponds to output labels.}}
            \label{fig:fig_attention}
        \end{figure}

Figure~\ref{fig:fig_attention} visualizes the self-attention alignments in the encoder layer and the vanilla attention alignments in the encoder-decoder layer by Tensorflow \cite{abadi2016tensorflow}. As can be seen in the figure, both self-attention matrix and vanilla attention matrix appear very localized, which let us to understand how changing the attention window influences the CER.

\subsection{Results of CI-phoneme and syllable based model}

Our results are summarized in Table~\ref{tab:subunit_word results}. As can be seen in the table, CI-phoneme and syllable based model with the Transformer can achieve competitive results on HKUST datasets in terms of CER. It reveals that the Transformer is very suitable for the ASR task since its powerful sequence modeling capability, although it relies entirely on self-attention without using RNNs or convolutions. Furthermore, we note here that the CER of syllable based model outperforms that of corresponding CI-phoneme based model. The results suggest that the sub-word unit of syllables is a better modeling unit in sequence-to-sequence attention-based models on Mandarin Chinese ASR tasks compared to the sub-word unit of  CI-phonemes. It validates the conclusion proposed on CTC based model \cite{qu2017syllable}.
Finally, it is obvious that the big model always performs better than the base model no matter on the CI-phoneme based model or syllable based model. Therefore, our further experiments are conducted on the big model.

We further generate more training data by linearly scaling the audio lengths by factors of $0.9$ and $1.1$ (speed perturb.) \cite{hori2017advances}.
It can be observed that syllable based model with speed perturb becomes better and achieves the best CER of \emph{$28.77\%$} compared to without it.
However, CI-phoneme based model with speed perturb becomes very slightly worse than without it. The interpretation of this phenomenon is that syllables have a longer duration and more invariance than CI-phonemes, so small speed perturb would not affect the pronuciation of syllables too much, instead of providing more useful and various training data. However, small speed perturb might have more impact on the pronuciation of CI-phonemes due to the short duration.

        \begin{table}[th]
        \caption{\label{tab:subunit_word results} {\it Comparison of CI-phoneme and syllable based model with the Transformer on HKUST datasets in CER (\%).}}
        \vspace{2mm}
        \centerline{
          \begin{tabular}{|c|c|c|}
            \hline
              {sub-word unit}   & {model}  &    {CER}   \\
            \hline
            \multirow{3} * {CI-phonemes} & D512-H8 &   $32.94$  \\
            & D1024-H16 &   \textbf{30.65}  \\
            & D1024-H16 (speed perturb) &   $30.72$  \\
            \cline{2-3}
            \hline
            \multirow{3} * {Syllables} & D512-H8 &   $31.80$  \\
            & D1024-H16 &   $29.87$  \\
            & D1024-H16 (speed perturb) &   \textbf{28.77}  \\
            \cline{2-3}
            \hline
          \end{tabular}
        }
        \end{table}


\subsection{Comparison with previous works}

In Table~\ref{tab:comparison_with_previous}, we compare our experimental results to other model architectures from the literature on HKUST datasets. First, we can find that the result of CI-phoneme based model with the Transformer is comparable to the best result by the deep multidimensional residual learning with 9 LSTM layers in hybrid system \cite{zhao2016multidimensional}, and the syllable based model with the Transformer provides over a $6\%$ relative improvement in CER compared to it.
Moreover, the CER \emph{$28.77\%$} of syllable based model with the Transformer is comparable to the CER \emph{$28.9\%$} by the joint CTC-attention based encoder-decoder network \cite{hori2017advances} when no external language model is used, but slightly worse than the CER \emph{$28.0\%$} by the joint CTC-attention based encoder-decoder network with separate RNN-LM, which is the state-of-the-art on HKUST datasets to the best of our knowledge.

      \begin{table}[th]
      \newcommand{\tabincell}[2]{\begin{tabular}{@{}#1@{}}#2\end{tabular}}
        \caption{\label{tab:comparison_with_previous} {\it CER (\%) on HKUST datasets compared to previous works.}}
        \vspace{2mm}
        \centerline{
          \begin{tabular}{|c|c|}
            \hline
              {model}   &    {CER}   \\
            \hline
            \tabincell{c}{LSTMP-9$\times$800P512-F444 \cite{zhao2016multidimensional}\ } &    $30.79$  \\
            \tabincell{c}{CTC-attention+joint dec. (speed perturb., one-pass) \\ +VGG net \\ +RNN-LM (separate) \cite{hori2017advances}\ } &    \tabincell{c}{ \\ $28.9$ \\ \textbf{28.0} }  \\
            \hline
             CI-phonemes-D1024-H16 &    $30.65$  \\
             Syllables-D1024-H16 (speed perturb) &    \textbf{28.77}  \\
            \hline
          \end{tabular}
        }
      \end{table}

\subsection{Comparison of different frame rates}

Finally, table~\ref{tab:transformer frames_results} compares different frame rates on CI-phoneme and syllable based model with the Transformer. It indicates that the performance of CI-phoneme and syllable based model with the Transformer decreases as the frame rate increases. The decreasing rate is relatively slow from $30$ms to $50$ms, but deteriorates rapidly from $50$ms to $70$ms. Thus, it shows that frame rate between $30$ms and $50$ms performs relatively well on CI-phoneme and syllable based model with the Transformer.

      \begin{table}[th]
      \newcommand{\tabincell}[2]{\begin{tabular}{@{}#1@{}}#2\end{tabular}}
        \caption{\label{tab:transformer frames_results} {\it Comparison of different frame rates on HKUST datasets in CER (\%).}}
        \vspace{2mm}
        \centerline{
          \begin{tabular}{|c|c|c|}
            \hline
              {model}   &    {frame rate}  &    {CER} \\
            \hline
            \tabincell{c}{CI-phonemes-D1024-H16 \\ (speed perturb)} &   \tabincell{c}{$30ms$ \\  $50ms$ \\ $70ms$} & \tabincell{c}{\textbf{30.72} \\ $31.68$ \\ $33.96$} \\
            \hline
            \tabincell{c}{Syllables-D1024-H16 \\ (speed perturb)} &   \tabincell{c}{$30ms$ \\ $50ms$ \\ $70ms$} & \tabincell{c}{\textbf{28.77} \\ $29.36$ \\ $32.22$} \\
            \hline
          \end{tabular}
        }
      \end{table}

\section{Conclusions}
\label{label_conclusions}

In this paper we applied the Transformer, a new sequence transduction model based entirely on self-attention without using RNNs or convolutions, to Mandarin Chinese ASR tasks and verified its effectiveness on HKUST datasets.
Furthermore, we compared syllables and CI-phonemes as the modeling unit in sequence-to-sequence attention-based models with the Transformer in Mandarin Chinese.
Our experimental results demonstrated that syllable based model with the Transformer performs better than CI-phoneme based counterpart on HKUST datasets.
What is more, a greedy cascading decoder with the Transformer is proposed to maximize $Pr(W|s)$$Pr(s|X)$ and then posterior probability $Pr(W|X)$ can be maximized.
Experimental results on CI-phoneme and syllable based model verified the effectiveness of the greedy cascading decoder.


\section{Acknowledgements}

The authors would like to thank Chunqi Wang for insightful discussions on training and tuning the Transformer.

\bibliographystyle{IEEEtran}

\bibliography{mybib}

\end{document}